\documentclass[doublecol,linenumbers]{epl2} 

\usepackage{enumerate}

\usepackage{amsmath}

\title{The influence of persuasion in opinion formation and polarization}
\shorttitle{Persuation and opinion polarization} 

\author{C. E. La Rocca\inst{1}\thanks{E-mail:
    \email{larocca@mdp.edu.ar}} \and L. A. Braunstein\inst{1,2} \and
  F. Vazquez\inst{3}\thanks{E-mail: \email{fede.vazmin@gmail.com}} }

\shortauthor{C. E. La Rocca \etal}

\institute{                    
  \inst{1} Instituto de Investigaciones
  F\'isicas de Mar del Plata, UNMDP-CONICET - 7600, Mar del
  Plata, Argentina\\
  \inst{2} Center for Polymer Studies, Physics Department, 
  Boston University - Boston, Massachusetts 02215, USA \\
  \inst{3} Instituto de F\'isica de L\'iquidos y Sistemas
  Biol\'ogicos, UNLP-CONICET - 1900, La Plata, Argentina 
}
\pacs{02.50.-r}{Probability theory, stochastic processes, and statistics}
\pacs{87.23.Ge}{Dynamics of social systems}
\pacs{05.10.-a}{Computational methods in statistical physics and
  nonlinear dynamics}

\abstract{We present a model that explores the influence of persuasion
  in a population of agents with positive and negative opinion
  orientations.  The opinion of each agent is represented by an integer
  number $k$ that expresses its level of agreement on a given issue, from
  totally against $k=-M$ to totally in favor $k=M$. 
  Same-orientation agents persuade each other  with probability $p$,
  becoming more extreme, while opposite-orientation agents  become
  more moderate as they reach a compromise with probability $q$.
  The population initially evolves to (a) a polarized state for
  $r=p/q>1$, where opinions' distribution is peaked at the
  extreme values $k=\pm M$, or (b) a centralized state for $r<1$, with
  most opinions around $k=\pm 1$.  When $r \gg 1$, polarization
  lasts for a time that diverges as $r^M \ln N$, where $N$
  is the population's size.   Finally, an extremist consensus 
  ($k=M$ or $-M$) is reached in a time that scales as 
  $r^{-1}$ for $r \ll 1$.}

\begin{document}

\maketitle

\section{Introduction}

Many empirical investigations show the importance of social influence
in the formation of people's opinions.  For instance, it is argued
that two interacting partners may exert social pressure to change
their attitudes to conform each other \cite{Festinger-1950}.  Some
physics models have incorporated this particular social mechanism by
means of a \emph{compromise process}
\cite{Castellano-2009,Weisbuch-2002,Ben-Naim-2003-1,Ben-Naim-2003-2}.  In these
models, opinions are represented by a real number between two extreme
values, and pair of individuals interact only if their opinion
difference is smaller than a given threshold.  Individuals resolve the
conflict by reaching a compromise, in which both opinions are changed
in the same amount to reduce their difference.  A less explored
mechanism of social interactions is the \emph{persuasive arguments
  exchange} \cite{Myers-1982,Isenberg-1986,Lau-1998,Mas-2012}.  As
observed by Myers \cite{Myers-1982} in group discussion experiments, when two
individuals talk, they do not only state their opinions, but they also
discuss about the arguments that support their opinions.  Then, if
they already hold the same opinion orientation, they could intensify
their opinions by persuading each other with new arguments or reasons, 
becoming more extreme in their believes.  This mechanism was proposed
by Lau and Murnighan \cite{Lau-1998} after the works by Myers
\cite{Myers-1982} and Isenberg \cite{Isenberg-1986}, and recently
explored by M\"as et al. \cite{Mas-2012} using a computational model.

\begin{figure}[ht]
\onefigure[width=0.45\textwidth]{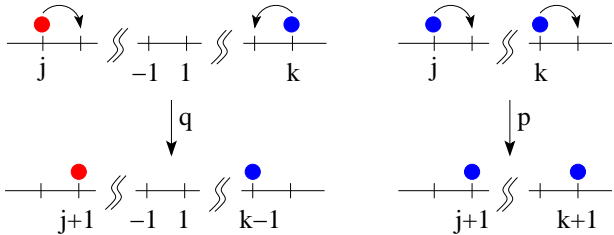}
\caption{Two main processes of the model: (Left) Compromise: two
  interacting agents with opposite opinion orientation become more
  moderate. (Right) Persuasion: two interacting agents with the same 
  orientation become more extremists.}
\label{c-p}
\end{figure}

In this letter, we introduce a simple model that explores the
competition between the compromise and persuasive-argument mechanisms
in a population of $N$ interacting agents.  The state of each agent is
represented by an integer number $k$ ($-M \le k \le M$ and $k \ne
0$), where the sign of $k$ indicates its opinion orientation, like for
instance to be in favor (positive) or against (negative) marijuana
legalization, and the absolute value $|k|$ measures its opinion
intensity or strength.  Thus, $k=M$ ($-M$) correspond to extremists
which are strongly in favor (against) of 
legalization, while $k=1$ and $-1$ represent moderates.  In a time
step, two agents with states $j$ and $k$ are picked at random to
interact.  Then, their states are updated according to two
elemental processes (see Fig.~\ref{c-p}). \\
(i) \emph{Compromise}: if they have opposite orientations, their intensities
decrease in one unit with probability $q$: 
\begin{itemize}
\item If $j<0$ and $k>0$ $\Rightarrow$ $(j,k) \to (j^r,k^l)$ with prob. $q$
\item If $j>0$ and $k<0$ $\Rightarrow$ $(j,k) \to (j^l,k^r)$ with
  prob. $q$.
\end{itemize}
If $j=\pm 1$ and $k=\mp 1$, one switches orientation at random:
\begin{eqnarray*}
(\pm 1, \mp 1) \to \left\{ \begin{array}{ll}
(1,1) & \mbox{with prob. $q/2$} \\   
(-1,-1) & \mbox{with prob. $q/2$.}\end{array}
\right.
\end{eqnarray*} 
(ii) \emph{Persuasion}: if they have the same orientation, their 
intensities increase by one unit with probability $p$:
\begin{itemize}
\item If $j<0$ and $k<0$ $\Rightarrow$ $(j,k) \to (j^l,k^l)$ with prob. $p$ 
\item If $j>0$ and $k>0$ $\Rightarrow$ $(j,k) \to (j^r,k^r)$ with prob. $p$. 
\end{itemize}
Here $k^r$ and $k^l$ denote the right and left neighboring states of
$k$, respectively, defined as
\begin{eqnarray*} 
k^r = \left\{ \begin{array}{lll}
1 & \mbox{for $k = -1$} \\   
M & \mbox{for $k=M$} \\
k+1 & \mbox{otherwise}.\end{array}
\right.~~~
k^l = \left\{ \begin{array}{lll}
-1 & \mbox{for $k = 1$} \\   
-M & \mbox{for $k=-M$} \\
k-1 & \mbox{otherwise}.\end{array}
\right.
\end{eqnarray*}

With this dynamics, opinions are constrained to the interval
$[-M,M]$ and the neutral opinion $k=0$ is excluded.  We find that the 
population's opinion settles in a centralized state
when the compromise process dominates ($q>p$), and in a polarized state when 
persuasion dominates ($p>q$).  These states are not stable, and the
system ultimately reaches extremist consensus.  We solve the equations
for the dynamics in the stationary state, and also in the strong and
small persuasion limits, and find that the mean extremist consensus
time is non-monotonic in the ratio $p/q$.   

We note that similar mechanisms to the compromise process (i) are
found in nonlinear and multiple-state voter models with a
reinforcement rule
\cite{Castello-2006,Dall'Asta-2007,Vazquez-2008,Castellano-2009-2,
Volovik-2012,Terranova-2014}, in which agents switch orientation 
(opinion's sign) only after receiving multiple inputs of agents with
the opposite orientation.
Besides, persuasion was used in recent works 
\cite{Crokidakis-2012,Crokidakis-2013} as a degree of a person's 
self-conviction, where in addition to the influence from others, a
person takes into account its own opinion when making a decision.
Also, persuasion between opposite-orientation agents was recently
studied in \cite{Terranova-2014}.  However, we understand that the mechanism
of strengthening of opinions due to same-orientation interactions has not been
investigated within an interacting particle model.    

\section{Dynamics}

\begin{figure}
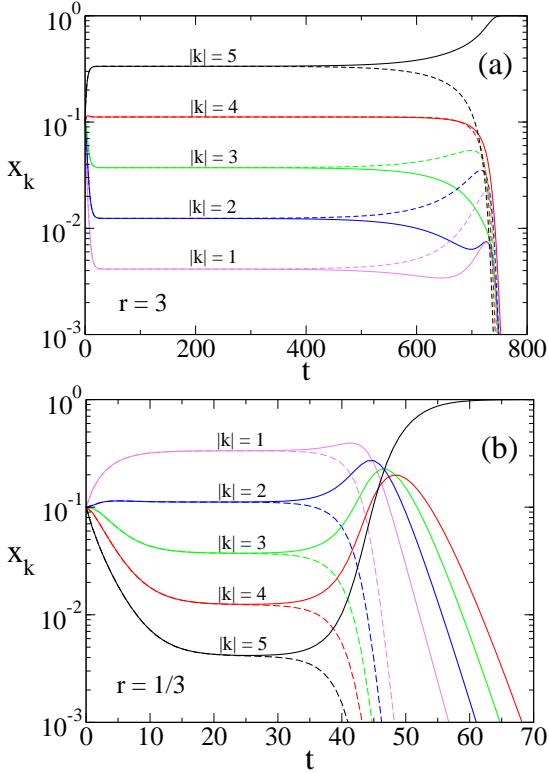

\onefigure[width=0.4\textwidth]{xn-t-MC-3.00.eps}
\onefigure[width=0.4\textwidth]{xn-t-MC-0.33.eps}
\caption{Time evolution of the fraction of agents in different opinion
  states, for maximum opinion intensity $M=5$ in a population of
  $N=10^9$ agents.  (a) $x_k(t)$ for $p=3/4$ and $q=1/4$. (b)
  $x_k(t)$ for $p=1/4$ and $q=3/4$.  Solid (dashed) curves correspond
  to positive (negative) opinions.  A logarithmic scale was used in
  the y-axis to clearly see all plateaus together.}
\label{xn-t}
\end{figure}

We study the dynamics of the system by looking at the time evolution
of the number of agents in the different opinion states.  We
denote by $x_k(t)$ the fraction of agents in state $k$ at time $t$.
Initially,  states are uniformly distributed, thus $x_k(t=0) \simeq
1/2M$.   Figure~\ref{xn-t} shows results from Monte Carlo (MC) simulations for
$M=5$ and a  population of size $N=10^9$.  Given that qualitative
results depend on the ratio $r \equiv p/q$ that relates the persuasion and
compromise time scales, we show two representative cases,
one with $r=3$ [Fig.~\ref{xn-t}(a)] and the other with $r=1/3$ 
[Fig.~\ref{xn-t}(b)].  We observe that densities $x_k$ reach a nearly 
constant value (plateau) that depends
on $k$, but eventually all $x_k$ decay to zero, except 
$x_{\mbox{\tiny M}}$ that goes 
to $1$, corresponding to a consensus in the extremist state $M$.  The
two extremists consensus $x_{\pm \mbox{\tiny M}}=1$ are absorbing states of the
system, thus they are the only possible final states in the long run.
The length of the plateau increases with the system size as $\ln N$
(not shown), a
typical time scale that appears in models with intermediate states
\cite{Castello-2006,Volovik-2012}.  We shall see that this
particular scaling is also a consequence
of the discrete nature of the system when a small initial asymmetry is
introduced \cite{Volovik-2012}.  

\begin{figure}[ht]
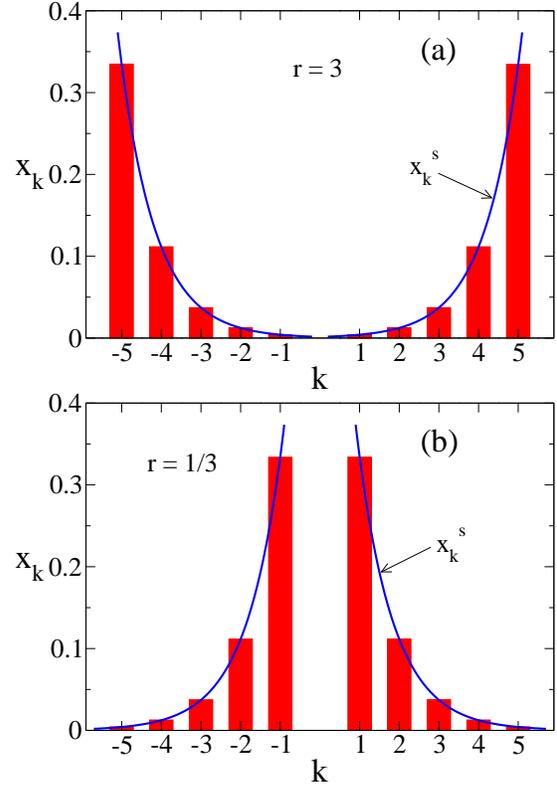

\onefigure[width=0.4\textwidth]{xn-s-MC-3.00.eps}
\onefigure[width=0.4\textwidth]{xn-s-MC-0.33.eps}
\caption{Distribution of opinions' densities at the quasistationary mixed state
  of Fig.~\ref{xn-t}, and for the same parameter values. (a) $x_k$ at time
  $t=200$. (b) $x_k$ at $t=20$.  Solid lines correspond to 
  expression~(\ref{xns}).}
\label{xn-s}
\end{figure}

The structure of the population at the quasistationary state or
plateau shows interesting properties, as can be seen in
Fig. \ref{xn-s} where we plot $x_k$ for a given time in the plateau.
The distribution of opinions depends on the ratio $r$, which
controls the relative frequency of  persuasion and compromise events.
When $r>1$, the persuasion process  dominates over compromise, driving
the states of agents towards the extreme opinions $k=\pm M$.  This
induces opinion polarization, where $x_k$  is symmetric and peaked at
the opposite extreme values [see Fig. \ref{xn-s}(a)].  Instead, for
$r<1$ compromise events occur more often than persuasive encounters,
thus most opinions accumulate around the moderate values $k=\pm 1$,
inducing a centralized opinion state where $x_k$ has a maximum value
at center states [see Fig. \ref{xn-s}(b)].  

\section{Stationary states}
\label{stationary}

To gain an insight about these observations, we write and analyze a set of 
ordinary differential equations for the time evolution of $x_k$.  Here we 
consider for simplicity the large $N$ limit, where demographic noise coming 
from system size fluctuations is neglected.  Then, the densities of 
positive states evolve according to the following set of equations
\begin{subequations}
\label{dxdt}
\begin{equation}
\label{dxdt1}
\hspace{0.5cm}\frac{d x_1}{d t} = 2\,(x_{-1}\,q-x_1\,p)\,\sigma_+
+2 q\,(x_{2}-x_1)\,\sigma_- 
\end{equation}
\begin{eqnarray}
\label{dxdt2}
\hspace{-1cm}\frac{d x_k}{d t} &=& 2p\,(x_{k-1}-x_k)\, \sigma_+ \\  
&+& 2q\,(x_{k+1}-x_k)\,\sigma_- ~~~\mbox{for}~~2 \le k \le M-1
\nonumber 
\end{eqnarray}
\begin{equation}
\label{dxdt3}
\hspace{-3.5cm}\frac{d x_{\mbox{\tiny M}}}{d t} = 
2p\, x_{\mbox{\tiny M-1}} \,\sigma_+ - 2q\, x_{\mbox{\tiny M}}\, \sigma_-,
\end{equation}
\end{subequations}
where $\sigma_+= \sum_{k=1}^Mx_k $ and 
$\sigma_-= \sum_{k=-1}^{-M}x_k$  are the total densities of positives
and negatives states, respectively, which satisfy the density conservation
constraint $\sigma_++\sigma_-=1$.  Equations for negative-state
densities are obtained from Eqs.~(\ref{dxdt}) by the transformations 
$k \leftrightarrow -k$ and $\sigma_+ \leftrightarrow \sigma_-$.  The
gain and loss terms in the
rate equations account for the different processes.  The first term
describes persuasive interactions between two positive agents, while
the second term accounts for the compromise between positive and negative
agents.  In addition, the gain term $2 q\, x_{-1}\,\sigma_+$ in 
Eq.~(\ref{dxdt1}) 
corresponding to $-1 \to 1$ transitions, describes the
negative to positive flux of states, while the absence of the loss
term $-2 p\, x_{\mbox{\tiny M}}\, \sigma_+$ and the gain term 
$2 q\, x_{\mbox{\tiny M+1}}\, \sigma_-$ in Eq.~(\ref{dxdt3}) 
reflect the fact that there is no state flux through the $k=M$ boundary. 
 
The properties of the quasistationary distributions of
Fig.~\ref{xn-s} can be obtained by studying the stationary solutions
of Eqs.~(\ref{dxdt}).  The two trivial
solutions $x_{\mbox{\tiny M}}=1$  and $x_{\mbox{\tiny -M}}=1$
correspond to the $M$ and $-M$
extremists consensus, respectively, where all agents end up with the
same maximum  opinion intensity.  These are stable fixed points in the
space of densities.  But there is also a non-trivial solution that
corresponds to a balanced mix of positive and negative agents, as the
ones in Fig.~\ref{xn-s}.  Setting
$\frac{dx_k}{dt}=0$ and $\sigma_+=\sigma_-=1/2$ in Eqs.~(\ref{dxdt})
we obtain a linear system of algebraic equations that can be solved by
iteration.  The solutions are $x_k^s=x_1^s\, r^{k-1}$ for $1 \le k \le
M$ and  $x_k^s=x_{-1}^s\, r^{-k-1}$ for $-M \le k \le -1$, with
$r=p/q$.  Using the
normalization condition  $1/2=\sigma_+=\sum_{k=1}^M
x_k=\frac{x_1(1-r^M)}{(1-r)}$ and $1/2=\sigma_-=\sum_{k=-1}^{-M}
x_k=\frac{x_{-1}(1-r^M)}{(1-r)}$, we obtain the values
\begin{equation}
x_1^s=x_{-1}^s=\frac{1-r}{2(1-r^M)}.
\label{x1s}
\end{equation}
Finally, densities at the quasistationary mixed state are
\begin{equation}
x_k^s = \frac{1}{2} \left( \frac{1-r}{1-r^M} \right) r^{|k|-1}
~~~\mbox{for}~~~ -M \le k \le M.
\label{xns}
\end{equation}
In Fig.~\ref{xn-s} we observe that expression (\ref{xns}) in solid
lines gives a good mathematical description of the opinions'
distributions from MC simulations, in a population of agents whose
opinions are polarized ($r>1$) or centralized ($r<1$).        

To study the stability of these states we have integrated
Eqs.~(\ref{dxdt}) numerically for $M=5$ and two values of $r$.  The
time evolution is very similar to the one depicted in Fig.~\ref{xn-t}.
We mimic the initial state of MC simulations by taking
$x_k(t=0)=1/2M+\epsilon$, where $|\epsilon|=N^{-1/2}$ corresponds to
a stochastic size fluctuation respect to the uniform state.  All
densities quickly reach a nearly constant value in time, corresponding to the
mixed solution $x_k^s$ of Eq.~(\ref{xns}), and stay very close to this
attractor for a time that scales as $\ln N$, to finally reach either fixed
point $x_{\pm \mbox{\tiny M}}=1$.  The attractor $x_k^s$ corresponds to a saddle
point of the dynamics - starting from the exact uniform state
$x_k(t=0)=1/2M$ (or any symmetric case $x_k=x_{-k}$) causes the system
to hit $x_k^s$, and stay there.  But any small initial
asymmetry, for instance in the positive opinion, makes the system stay
in the vicinity of $x_k^s$ for a finite time, and
eventually escape and hit the positive extremist consensus state
$x_{\mbox{\tiny M}}=1$.  The time spent near the saddle point is
related to the time to reach a consensus in orientation (all
states with the same sign) and, as we show in the next section, is 
non-monotonic in $r$.

\section{Convergence times}
\label{Convergence}

\begin{figure}[ht]
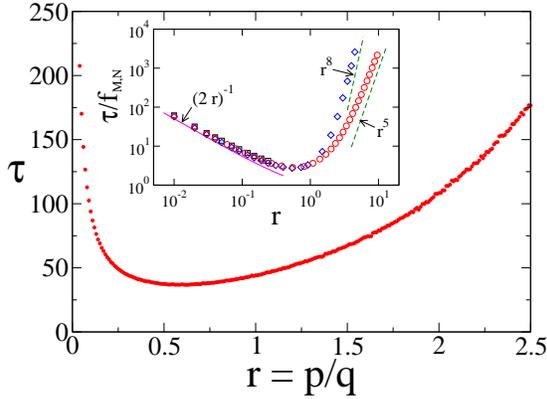

\onefigure[width=0.4\textwidth]{tau-r.eps}
\caption{Main: mean extremist consensus time $\tau$ vs persuasion strength
  $r=p/(1-p)$ for $M=5$ and $N=1000$.  Inset: rescaled time $\tau$ vs
  $r$ in log-log scale for $M=3$ (squares), $M=5$ (circles) and $M=8$
  (diamonds).  The solid line corresponds to the approximation 
  (\ref{tau-r-N-M}) in the $r \ll 1$ limit, while dashed lines denote
  the asymptotic behavior $r^M$ in the $r \gg 1$ limit.}
\label{tau-r}
\end{figure}

In Fig.~\ref{tau-r} we plot the mean time $\tau$ to reach the final extremist
consensus $x_{\pm \mbox{\tiny M}}=1$ as a function of $r$ for
$M=5$, obtained from MC simulations.  As qualitative
results only  depend on $r$ we took $q=1-p$, thus $r=p/(1-p)$ varies
from $0$ to $\infty$ as $p$ goes from $0$ to $1$.  Therefore, $r$ can
be seen as the relative strength of persuasion, as compared to
compromise.  We observe that $\tau$ is non-monotonic in $r$, and has a
minimum value around $r \simeq 0.6$.  This means that the population reaches
the fastest consensus when interactions between agents of the same
orientation have a probability of success $p$ similar to that of
opposite-orientation agents $q$.  Instead, mostly chatting with
same-opinion partners (large $r$) reinforce initial believes, leading
to a polarized state that last for very long times.  Besides, only
interacting with opposite-opinion partners (small $r$) first induces a
centralized consensus, which is unstable, and then the population is
slowly driven to the final extremist consensus.    

An insight about the non-monotonic behavior of $\tau$ can be obtained
by means of Eqs.~(\ref{dxdt}).   
For a simpler analysis of the equations and a better understanding of
the previous results, it proves convenient to
split the evolution of the system into two distinct stages - a first
stage with an associated time scale $\tau_1$, in which all agents
adopt the same opinion-orientation (all states are either positive or
negative), and a second stage where the
system reaches extremist consensus, characterized by a time
scale $\tau_2$.  Therefore, the convergence time can be written as 
$\tau=\tau_1+\tau_2$.  The non-linearity of Eqs.~(\ref{dxdt}) makes
it hard to find a complete solution, but it is possible to obtain
approximate expressions for $\tau$ in the two limiting cases of very
strong and very weak persuasion.

\subsection{Small persuasion limit $r \ll 1$}

In this limit, the second stage is much longer
than the first stage ($\tau_2 \gg \tau_1)$, and we can approximate
$\tau \simeq \tau_2$.  This is because once all agents' states become positive
(negative) they are slowly driven by persuasion events - which happen with
a very small probability $p=r/(1+r)$ - to
the consensus state $x_{\mbox{\tiny M}}=1$ ($x_{\mbox{\tiny -M}}=1$),
thus the system spends most of 
the time in the second stage.  To estimate $\tau_2$ we assume, without
loss of generality, that the system starts at time $t=0$ from a
configuration in which all states are positive 
[$x_k(t=0)=0 ~~\forall\, k<0$]. 
This initial condition implies that states remain positive since only
persuasive events can take place, and thus $\sigma_+(t)=1$ and
$\sigma_-(t)=0$ for $t \ge 0$.  
Then, Eqs.~(\ref{dxdt}) become linear
\begin{eqnarray}
\label{dxdt-p1}
\frac{dx_1}{dt'} &=& -x_1 \nonumber \\ 
\frac{dx_k}{dt'} &=& x_{k-1}-x_{k}~~~\mbox{for}~~~2 \le k \le M-1 \\
\frac{dx_{\mbox{\tiny M}}}{dt'} &=& x_{\mbox{\tiny M-1}}  \nonumber,
\end{eqnarray}
where we have introduced the rescaled time $t'\equiv 2 \, p \, t$.  In the
Laplace space, Eqs.~(\ref{dxdt-p1}) are reduced to the following system of
coupled algebraic equations:
\begin{eqnarray*}
s \, x_1(s) - x_1(0) &=& - x_1(s) \nonumber \\ 
s \, x_k(s) - x_k(0) &=&  x_{k-1}(s)-x_{k}(s)~~~2 \le k \le M-1 \nonumber \\
s \, x_{\mbox{\tiny M}}(s) - x_{\mbox{\tiny M}}(0) &=&  
x_{\mbox{\tiny M-1}}(s),
\end{eqnarray*}
with solutions 
\begin{eqnarray*}
x_k(s) &=& \sum_{n=0}^{k-1} \frac{x_{k-n}(0)}{(s+1)^{n+1}} 
~~~\mbox{for}~~~1 \le k \le M-1 \\\\
x_{\mbox{\tiny M}}(s) &=& \frac{x_{\mbox{\tiny M}}(0)}{s} +
\frac{1}{s} \, \sum_{n=0}^{M-2} \frac{x_{k-n}(0)}{(s+1)^{n+1}}. 
\end{eqnarray*}
Transforming back to the original space and replacing $t'$ by 
$2\, p \, t$ we finally obtain
\begin{eqnarray} 
x_k(t) &=& e^{-2p\,t} \, \sum_{n=0}^{k-1} \frac{(2 p\,t)^n \,
x_{k-n}(0)}{n!}~~~1 \le k \le M-1 \nonumber \\ \nonumber \\
x_{\mbox{\tiny M}}(t) &=& 1 - e^{-2p\,t} \, \sum_{k=1}^{M-1} \sum_{n=0}^{k-1}
\frac{(2p\,t)^n \, x_{k-n}(0)}{n!}.
\label{xM-t}
\end{eqnarray}
The above solutions are valid for all values of $r$, but we explore
here their behavior in the $r \ll 1$ limit.  In this case we expect an
initial distribution of
states peaked at $k=1$, that is, $x_1(0) \simeq 1$ and 
$x_k(0) \simeq 0$ for $2 \le k \le M$. This is because the strong bias
towards the
center during the first stage keeps most states close to $k=1$.  Then,
Eq.~(\ref{xM-t}) becomes
\begin{equation}
x_{\mbox{\tiny M}}(t) \simeq 1 - e^{-2p\,t} \, \sum_{k=0}^{M-2} \frac{(2p\,t)^k}{k!},
\label{xM-t2}
\end{equation}
which shows that $x_{\mbox{\tiny M}}$ approaches $1$ quasi-exponentially fast with
time.  Having $x_{\mbox{\tiny M}} > 1-1/N$ at a time $t=\tau_2$ is
equivalent to an extremist consensus in the discrete
system of $N$ agents, since this corresponds to have a number of
agents in state $M$ larger than $N-1$.  Therefore, from
Eq.~(\ref{xM-t2}) $\tau_2$ obeys the following relation 
\begin{equation}
N\, e^{-2p\,\tau_2} \sum_{k=0}^{M-2} \frac{(2p\,\tau_2)^k}{k!}=1 
\label{xM-t4}.
\end{equation}
Then, $\tau_2 = f_{\mbox{\tiny M,N}}/2p$, where $f_{\mbox{\tiny M,N}}$
is a solution of 
\begin{equation}
N e^{-f} \sum_{k=0}^{M-2} f^k/k!-1 = 0,
\label{xM-t5}
\end{equation}
a non-trivial function of $M$ and $N$.  
Finally, replacing back $p = r/(1+r)$ we arrive to following
expression for $\tau$
\begin{equation}
\tau \simeq \tau_2 \simeq \frac{(1+r) f_{\mbox{\tiny M,N}}}{2r}.
\label{tau-r-N-M}
\end{equation}
In the inset of
Fig.~\ref{tau-r} we show the curves $\tau$ vs $r$ from MC simulations
in a system of size $N=1000$, and rescaled by the 
functions $f \simeq 9.233, 13.062$ and $18.062$, for $M=3,5$ and $8$,
respectively.  These values of $f$ were obtained by numerically
solving Eq.~(\ref{xM-t5}), given that a closed expression
for $f$ in terms of $M$ and $N$ is very hard to obtain.  
The collapse of the three curves confirm the scaling given by
Eq.~(\ref{tau-r-N-M}), which also captures the $r \to 0$ asymptotic
behavior $r^{-1}$ observed from simulations.

\subsection{Large persuasion $r \gg 1$ limit}

In this case, the first stage takes much longer than the second stage,
and thus $\tau \simeq \tau_1$.  The system quickly becomes polarized by the
driving bias towards the extreme states $k=\pm M$, and stays polarized
for very long times, given that the flux of particles from one side to
the other is limited by the very small compromise probability
$q=1/(1+r)$.  To estimate $\tau_1$, it proves useful to work with
the magnetization $m$, defined as the difference between the fraction
of positive and negative states  
\begin{equation}
m(t) \equiv \sigma_+(t)-\sigma_-(t) = 2 \sum_{k=1}^M x_k(t) - 1.  
\end{equation}
From Eqs.~(\ref{dxdt}), the magnetization evolves according to 
\begin{equation}
\frac{d m}{dt}= 4 \,q \,(x_{-1} \sigma_+ - x_1 \sigma_- ),
\label{dmdt-0}
\end{equation}
or, using the relations $\sigma_\pm=(1 \pm m)/2$, is  
\begin{equation}
\frac{d m}{dt}= 2 \, q \left[x_{-1} (1+m) - x_1 (1-m) \right].
\label{dmdt}
\end{equation}
Equation~(\ref{dmdt-0}) can also be obtained by noting that $m$ only changes
after a compromise event that involves states $1$ or $-1$.  The first term
accounts for $-1 \to 1$ transitions due to compromises between agents
with states $-1$ and $k>0$, which happen at a rate $2\, x_{-1}\,
\sigma_+$, increasing $m$ by $2/N$.  The second term stems for the
reverse transition $1 \to -1$, where $m$ decreases.
Equation~(\ref{dmdt}) is not closed because $x_{\pm 1}$ depend on
$x_{\pm 2}$, which in turn depend on $x_{\pm 3}$ and so on, as we
observe from Eqs.~(\ref{dxdt}).  However, we can still close the equation by
finding approximate expressions for $x_{\pm 1}$ in terms of $m$, as we
detail below.  As we showed before, the distribution of opinions at
the quasistationary mixed state follows the
exponential relation $x_{\pm k}^s = x_{\pm 1}^s \, r^{k-1}$ ($1 \le k \le
M$).  Monte Carlo simulations show that the distribution remains 
exponential during
the first stage, $x_{\pm k}(t) = x_{\pm 1}(t) \, \alpha_\pm^{k-1}(t)$,
where $\alpha_\pm(t)$ are time-dependent variables.  Interestingly, we
have numerically checked that $\alpha_\pm(t)$ are
almost constant over time, and only a significant change is
observed at the very end of the stage.  Therefore, they can be considered as
slow variables, as compared to $m$, and taken as constants and equal
to their initial values $\alpha_\pm(t) \simeq \alpha_\pm(0)$. Thus, we can write
\begin{eqnarray*}
\sigma_\pm=\frac{1 \pm m}{2} \simeq x_{\pm 1}(t) \sum_{k=1}^M
\alpha_{\pm}^{k-1}(0) =  
x_{\pm 1}(t) \left[ \frac{1-\alpha_{\pm}^M(0)}{1-\alpha_{\pm}(0)} \right], 
\end{eqnarray*}
from where
\begin{equation}
x_{\pm 1}(t) \simeq \frac{[1-\alpha_\pm(0)]}{2[1-\alpha_\pm^M(0)]}  
[1 \pm m(t)].
\label{x1-t}
\end{equation}
Given that the quasistationary state is reached in a fast time scale
that is $\mathcal O(1)$ [see Fig.~\ref{xn-t}(a)], we neglect this
short transient and assume that the initial condition corresponds to
the stationary solution Eq.~(\ref{xns}).  Therefore, from Eq.~(\ref{x1-t}),
the initial variables $\alpha_{\pm}(0)$ obey 
\begin{eqnarray}
\frac{[1-\alpha_\pm(0)]}{2[1-\alpha_\pm^M(0)]} \simeq \frac{x_1^s}{1 \pm m_0},
\label{x1-t-2}
\end{eqnarray}
where $m_0=m(0)$ is the initial magnetization, and
$x_1^s=(1-r)/2(1-r^M)$ is the state-$1$ density at the quasistationary
state [Eq.~(\ref{x1s})].  Note that starting from the perfectly
symmetric mixed state gives $m_0=0$, and thus $\alpha_\pm(0)=r$.  
From Eqs.~(\ref{x1-t}) and (\ref{x1-t-2}) we get
\begin{eqnarray}
x_{\pm 1}(t) \simeq \frac{x_1^s}{1 \pm m_0} [1 \pm m(t)].  
\label{x1-t-3}
\end{eqnarray}
Plugging this expression for $x_{\pm 1}$ into Eq.~(\ref{dmdt}) leads to 
\begin{equation}
\frac{d m(t)}{dt} \simeq \frac{4 \, q \, x_1^s \, m_0}{1-m_0^2} [1-m(t)^2].
\label{dmdt-2}
\end{equation}
The integration of Eq.~(\ref{dmdt-2}) gives
\begin{equation}
m(t) \simeq \frac{(1+m_0) e^{A t} - (1 - m_0) e^{-A t}}{(1+m_0) 
e^{A t} + (1-m_0) e^{-A t}},
\label{m-t}
\end{equation}
where 
\begin{equation}
A \equiv \frac{4 \, q \, x_1^s \, m_0}{1-m_0^2} = 
\frac{2(1-r) m_0}{(1+r)(1-r^M)(1-m_0^2)}
\end{equation}
is the prefactor of Eq.~(\ref{dmdt-2}).  Expression~(\ref{m-t}) captures the
qualitative behavior of the magnetization, which approaches to
$|m|=1$ as 
\begin{eqnarray}
|m(t)| \simeq 1 - \frac{2(1-|m_0|)}{(1+|m_0|)} e^{-2|A|t}.
\label{m-t-2}
\end{eqnarray}
Within this framework of rate equations, the first stage ends at a time
$\tau_1$ when $|m|$ equals $1-1/N$, that is, when less than
one particle remains in one of the two sides.  From Eq.~(\ref{m-t-2}) we obtain
\begin{equation}
\tau \simeq \tau_1 \simeq \frac{(1-m_0^2)(1+r)(1-r^M)}{4\,|m_0|\,(1-r)}
\ln \left[\frac{2 N (1-|m_0|)}{1+|m_0|} \right].
\end{equation}
The scaling $\tau \sim r^M$ gives the right asymptotic behavior for $r
\gg 1$ (inset of Fig.~\ref{tau-r}).

\section{Summary and Conclusions}

In summary, we proposed and studied a model that incorporates two
mechanisms for the formation of opinions -
compromise and persuasion.   Compromise interactions between
individuals tend to moderate their opinions, while
persuasive contacts lead to extreme positions.  When compromise events
are more frequent than persuasive events, opinions are grouped
around moderate values, leading to a centralized state of opinions.
In the opposite case, if persuasion events dominate over compromise
events, opinions are driven towards extreme positive and negative
values, inducing polarization.  The centralized and
polarized states are unstable, and consensus in either positive or
negative extreme opinions is eventually achieved.  For a symmetric
initial distribution of opinions, these final extremist states are
equiprobable, but any asymmetry in the initial condition that favors a
given opinion orientation makes the population reach consensus in the
extreme state of the favored orientation.  The mean extremist
consensus time $\tau$ is
non-monotonic in the ratio $r=p/q$ between the probabilities of
successful persuasive and compromise events, and has a minimum when $p$
and $q$ are of the same order of magnitude.  In the small ($r \ll 1$) and large
($r \gg 1$) persuasion limit, the consensus time scales as 
$\tau \sim r^{-1}$ and $\tau \sim r^M \ln N$,
respectively, with the maximum intensity $M$ and population size $N$.

In the studied model, individuals reinforce their opinions by talking
to other partners with the same opinion orientation.  It would worth
while to explore some extensions that include a reinforcement
mechanism between individuals with opposite orientations.  Related to
that, it was recently found that a rejection rule between very
dissimilar individuals enhances polarization \cite{Chau-2013}.

\acknowledgments

We acknowledge financial support from grant FONCyT (Pict 0293/2008).

\end{document}